\documentclass[aps,pre,amsmath,amsfonts,amssymb,floatfix,superscriptaddress,nofootinbib,showkeys]{revtex4-1}
\pdfoutput=1
\setcitestyle{authoryear,round}
\usepackage{graphicx}  
\usepackage{dcolumn}   
\usepackage{bm}        
\usepackage{amsmath,amsthm,amscd,amsfonts,amssymb,hyperref,floatrow}
\usepackage{color}
\usepackage{bbm}
\usepackage{enumerate}
\usepackage[normalem]{ulem}

\newcommand{\E}[1]{\mathbb{E}[{#1}]}
\newcommand\be{\begin{equation}}
\newcommand\bea{\begin{eqnarray} \nonumber }
\newcommand\ee{\end{equation}}
\newcommand\eea{\end{eqnarray}}

\def\d{\mathrm d}
\def\D{\mathcal D}
\def\L{\mathcal L}
\def \E{ \mathbb E  }
\begin{document}

\title{From Walras' auctioneer to continuous time double auctions\\
\small{A general dynamic theory of supply and demand}}

\author{J.~Donier} \email{jonathan.donier@polytechnique.org} \affiliation{Capital Fund Management, 23-25 Rue de l'Universit\'e, 75007 Paris, France.}\affiliation{Laboratoire de Probabilit\'es et Mod\`eles Al\'eatoires, Universit\'e Pierre et Marie Curie (Paris 6).}
\author{J.-P.~Bouchaud}\email{jean-philippe.bouchaud@cfm.fr} \affiliation{Capital Fund Management, 23-25 Rue de l'Universit\'e, 75007 Paris, France.}\affiliation{CFM-Imperial Institute of Quantitative Finance, Department of Mathematics, Imperial College, 
180 Queen's Gate, London SW7 2RH}
\date{\today}


\begin{abstract}
In standard Walrasian auctions, the price of a good is defined as the point where the supply and demand curves intersect. Since both curves are generically regular, the response to small perturbations is linearly small. However, a crucial ingredient is absent of the theory, namely transactions themselves. What happens after they occur? To answer the question, we develop a dynamic theory for supply and demand based on agents with heterogeneous beliefs. 
When the inter-auction time is infinitely long, the Walrasian mechanism is recovered. When transactions are allowed to happen in continuous time, a peculiar property emerges: close to the price, supply and demand vanish quadratically, which we empirically confirm on the Bitcoin. This explains why price impact in financial markets is universally observed to behave as the square root of the excess volume. The consequences are important, as they imply that the very fact of clearing the market makes prices hypersensitive to small fluctuations.
\end{abstract}

\maketitle

\section{Introduction}

One of the most time-worn statement of economic science is that ``prices are such that supply matches demand''. In order to explain how this really comes about, one 
usually invokes a Walras auctioneer, who attempts to measure the supply and demand curves $S(p)$ and $D(p)$, that give the total amount of supply/demand for a 
given good (or asset), would the price be set to $p$. The equilibrium price $p^*$ is then such that $D(p^*) = S(p^*)$, which maximizes the amount of good exchanged among agents, given the set of preferences corresponding to the current supply and demand curves \cite{walras2013elements}. In reality, the full knowledge of $S(p)$ and $D(p)$ is problematic, and Walras envisioned his famous {\it t\^atonnement} process as a mean to observe the supply/demand curves.  However, there is a whole aspect of the dynamics of markets that is totally absent in Walras' framework. While it describes how a pre-existing supply and demand would result in a clearing price, it does not tell
us anything about what happens \emph{after} the transaction has taken place. In this sense, the Walrasian price is of very limited scope, since the theory ceases to apply as soon as the price is discovered.

A practical solution to match supply and demand is the so-called ``order book'' \cite{harris1990liquidity, glosten1994electronic}, where each agent posts the quantities s/he is willing to buy or sell as a function of the price $p$. $S(p)$ (resp. $D(p)$) is then the sum of all sell (buy) quantities posted at or above (below) price $p$. At each time step, the auctioneer 
can then clear the market by finding the (unique) price such that $D(p^*) = S(p^*)$. This is 
in fact how most financial markets worked before the advent of electronic matching engines, when market makers played the role of ``active'' Walrasian auctioneers, in the sense that they would themselves contribute to the order book as to insure orderly trading and stable prices  \cite{glosten1985bid,madhavan2000market}.

Although close to Walras' idealization, order book based auctions are still confronted with a fundamental problem: agents do not necessarily reveal their intentions by placing visible orders, for fear of giving away information to the rest of the market -- among other reasons  \cite{handa1996limit,bongiovanni2006let}. It is plausible that only agents with the most urgent need to buy or to sell reveal their intentions. Only close to the transaction price is the order book expected to reveal the true underlying supply and demand curves $S(p)$ and $D(p)$, where they however get intertwined with the orders of market makers/high frequency traders who play strategic ``hide and seek'' games  \cite{handa2003quote,rosu2006multi,foucault2013market,bouchaud2008markets}. The visible order book is a sort of \emph{Potemkin village} that reveals only very little about the true underlying supply and demand\footnote{
The overwhelming activity market makers/HFT in the visible order book is a distraction from our story here:  since their position quickly mean-reverts around zero, 
they chiefly act as intermediaries (as they should) and  only weakly contribute to the supply and demand curves. See the detailed discussion in
 \cite{donier2014fully} and below.} and whose features strongly depend to the precise design of the market (time priority, pro-rata matching, small or large tick, presence of hidden orders, etc. -- see e.g. \cite{kockelkoren2010order}). A direct empirical observations of the dynamics of the full supply and demand curves $S(p)$ and $D(p)$ is therefore difficult (except in particular markets such as Bitcoin, see below and  \cite{donier2015markets}). But since the dynamics of prices is essentially governed by that of supply and demand, we need a plausible theoretical framework to model the (unobservable) evolution of the time dependent curves $S(p,t)$ and $D(p,t)$, where $t$ is time, to account for the (observable) evolution of prices. This would allow one to construct a ``Walrasian'' description of market dynamics, offering a much deeper level of understanding than simply postulating ad-hoc stochastic models for prices, such as the standard (geometric) Brownian motion  \cite{bachelier1900theorie,black1973pricing}. 

There are indeed many questions, some of utmost fundamental and practical importance, which cannot be addressed within these stochastic models and require the knowledge of the underlying supply and demand structure and dynamics. One of them is {\it price impact} \cite{bouchaud2010price}, i.e. how much does an additional unconditional buy/sell quantity $Q$ 
move the price up/down? This is important both for practitioners who want to estimate the costs associated to the impact of their trading strategies \cite{almgren2001optimal}, and for
regulators who want to understand the stability of markets and the price sensitivity to large ``freak'' orders (see e.g.  \cite{donier2015markets}). It is also of interest for the general understanding of price discovery and market efficiency: how much noise do ``noise traders'' \cite{kyle1985continuous} introduce in markets through their impact on prices? How relevant is marked to market accounting?  \cite{amihud1986asset,caccioli2012impact}, etc.

In a Walrasian context, the impact ${\cal I}$ of a small buy quantity $Q$ is easily shown to be linear in $Q$, simply because the slopes of the supply and demand curves around the price $p^*$ (that would prevail for $Q=0$) are generically non zero. More precisely, writing that $S(p_Q) = D(p_Q) + Q$ and Taylor expanding 
$S(p)$ and $D(p)$ around $p^*$ to first order in $Q$, one readily obtains:\footnote{The notation $\partial_p$ means that one takes the derivative with respect to the price $p$. More generally, $\partial_x$ means the derivative with respect to any variable $x$.}   
\be \label{Kyle-lambda1}
S(p^*) + (p_Q - p^*)\partial_p S(p^*) = D(p^*) + (p_Q - p^*)\partial_p D(p^*) + Q \qquad \Rightarrow \qquad {\cal I}(Q)\equiv  p_Q - p^* = \lambda Q
\ee
with $\lambda^{-1} = \partial_p S(p^*) - \partial_p D(p^*) > 0$, since one expects $S(p)$ to be a strictly increasing function of $p$ and $D(p)$ a strictly 
decreasing function of $p$ (see Fig. 1). Whenever the derivatives of the supply and demand curves do not simultaneously vanish at $p^*$, the price response to a perturbation must be {\it linear}. This intuitive result can also be justified using much more elaborate arguments, such as provided by
the Kyle model \cite{kyle1985continuous}, where noise traders, market makers and an informed trader interact in the market place. As shown by Kyle, the optimal strategy of market makers is to shift the price linearly in the market imbalance, with the coefficient $\lambda$ (``Kyle's lambda'') proportional to the volatility of the asset $\sigma$ and inversely proportional to the typical volume $V$ traded by the whole market. 

\begin{figure}[h]
    \begin{center}
        \includegraphics[height=6cm]{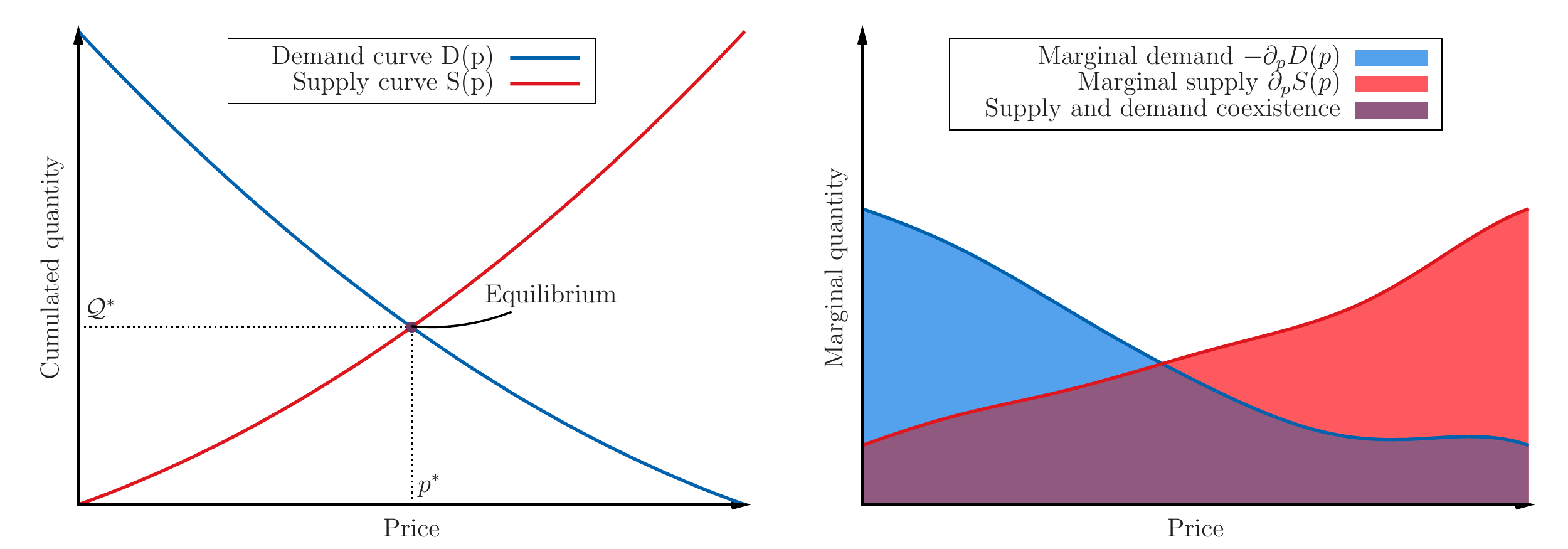}
    \end{center}
    \caption{Left: Illustration of the supply and demand curve (SD), and the resulting price according to Walras' law. 
    Right: Marginal supply and demand curves (MSD) corresponding to the figure on the left.}
    \label{fig:SD}
\end{figure}

Until very recently, the above linear relation between order imbalance and price changes was taken for granted in most academic papers. Remarkably, a series of independent empirical studies of the impact of proprietary orders in financial markets published since the mid-nineties suggests otherwise 
 \cite{Barra:1997,grinold2000active,Almgren:2005,Moro:2009,Toth:2011,Iacopo:2013,Gomes:2013,Bershova2013,Brokmann:2014,bacry2014market,donier2014million}. All these studies 
report a {\it strongly concave} price impact, even in the regime where $Q/V$ is very small (say between $10^{-4}$ and $0.1$). In fact, a simple square-root law
\be\label{sqrt_law}
\mathcal{I}(Q) = Y\sigma \sqrt{\frac{Q}{V}}
\ee
where $Y$ is a constant of order unity, accounts surprisingly well for a number of observations, quite independently of the type of markets (stocks, futures, 
FX, options,..), geographical zones, epochs (pre-2005, before the advent of massive HFT, or post-2005), trading style, etc. The empirical evidence is now 
so compelling -- see in particular the Bitcoin data in  \cite{donier2014million}  -- that it is difficult to avoid looking for a consistent theoretical explanation for such a universal non-linear impact law. 

There are actually two reasons to believe that usual equilibrium models are not relevant to explain the observed square-root impact. The first one is about 
orders of magnitude in the Bitcoin case, where the square-root impact is perfectly obeyed for price changes 30 times smaller than transaction fees 
themselves, and 300 times smaller than the daily volatility -- see \cite{donier2014million}. Imagining that the typical amateur trader on Bitcoin is able to
optimize anything with this level of precision seems a total utopia. The second reason is that a square-root impact formally corresponds to $\partial_Q {\cal I}(0) = \lambda \to \infty$, i.e. a situation where $\partial_p S(p^*)$ and $\partial_p D(p^*)$ are both zero. This is a fact that equilibrium models (as in \cite{kyle1985continuous}) cannot reproduce, as they would always predict a linear impact for small quantities.

A more likely possibility is that this square-root law is not willingly enforced by any market participant but is rather an emergent property. Stepping away from classical ideas, a detailed scenario for the divergence of 
Kyle's $\lambda$ was proposed in \cite{Toth:2011}. The main assumption in that paper is that the shape of the supply and demand curves is 
primarily the result of the interaction between order flow and past transactions themselves. This is in line with the idea that large 
population of interacting agents, each using heuristic decision rules, can lead to universal emergent behaviour (see e.g.  \cite{hommes2006heterogeneous} for a review of heterogeneous agent models in economics and finance \cite{gualdi2015tipping} for a recent 
didactic discussion). This scenario in fact emphasizes the \emph{transient} aspects of market dynamics, which are usually discarded in equilibrium models. The resulting detailed theory, elaborated in  \cite{Iacopo:2014,donier2014fully}, accounts well for the above square-root impact law and for many other empirical observations.  In
fact, the model presented in \cite{Toth:2011,donier2014fully} and in the present paper is not incompatible with game theoretic approaches as it can be rephrased in the language of the mean field games 
introduced in \cite{lasry2007mean}, see below, Sec. \ref{sec:agent-opt}. 

The aim of the present paper is to revisit and extend the Walrasian theory by proposing a dynamic theory of supply and demand in the light of these recent results. After a brief review of the literature in Sec. \ref{sec:context}, we propose in Sec. \ref{sec:dynamics} a dynamical theory for the evolution of the supply and demand curves $S(p,t)$ and $D(p,t)$, including transactions. Our theory only relies on weak, general hypotheses about the behaviour of agents, in particular the assumption of heterogeneous beliefs in a very large population. The Walrasian auctions setting can be seen as a limiting case of our theory, corresponding to an infinitely long time between auctions. We then show (Sec. \ref{sec:theory}) that as soon as the inter-auction time $\tau$ is finite, the impact of small volumes is linear, but with a coefficient $\lambda$ that diverges as $1/\sqrt{\tau}$ in the $\tau \to 0$ limit (corresponding to continuous time double auctions) where we recover an exact square-root impact.
This reflects the fact that the supply and demand curves both vanish quadratically close to the current price $p^*$, a property that we 
validate empirically using Bitcoin data.
For small, but finite $\tau$, impact is linear for very small $Q$s and becomes square-root beyond a crossover value $Q^*(\tau) \propto \tau$ when $\tau \to 0$.
Finally, in  Secs. \ref{sec:discussion},\ref{sec:conclusion} we
discuss some of the conceptual aspects of our framework, as well as some insights concerning market design and market stability, in particular in view of the recent proposals to curb the 
HFT activity by reintroducing periodic batch auctions \cite{budish2013high,fricke2014liquidity}. 

\section{Review of the literature}\label{sec:context}

The literature on price formation is obviously old and vast. It is divided into two distinct branches: microeconomics and financial economics, with quite different perspectives on the problem. The microeconomic community 
often attempts to determine how an economy with a pre-defined set of agents and preferences produces equilibrium prices, and study their properties (uniqueness, stability, computability, convergence etc.). This is 
mostly a static view, whereas financial economists are mostly interested in the dynamics of these prices. However, the assumption that markets are instantaneously arbitraged and efficient imposes (semi-)martingale 
properties for the price, the dynamics of which is entirely driven by \emph{news}, subsuming all knowledge of the actual dynamics of supply and demand itself. The quantities of interest are then the volatility of prices, 
the distribution of returns, etc., as well as the micro-structural properties of the immediate supply and demand visible in the order book, which however corresponds to an infinitesimal fraction of the total supply and
demand (see below). The aim of this section is to give a (rough) overview of these two different approaches and position the present work with respect to both of them.

\subsection{Theory of supply and demand in economics}\label{sec:eco}

As stated above, the question of how prices emerge from supply and demand has fuelled more than a century of economic research, based on the assertion that prices are such that supply equals demand for every asset in the economy. 
The immediate questions that arise are whether this equilibrium exists, is unique and is stable (in some sense) \cite{mas1995microeconomic}. Whereas these questions are rather subtle in a multi-asset economy  \cite{hicks1946value,samuelson1983foundations}, they become trivial in the case of a single asset economy as soon as the supply and demand curves are strictly monotonous. We will only consider a single asset economy in the present paper, since it fully suffices for our purpose and allows one to better focus on the essential part of our message, as we skip some of the usual problems that arise in multi-dimensional settings 
(is the equilibrium unique, stable, computable, etc.).

As many have noted, such a static description of prices is not fully satisfying, and a description of the dynamics of prices would be highly valuable. However, several interpretations of what ``dynamics" actually means 
can be found in the literature. 
\begin{enumerate}
\item Dynamics might refer to the way prices converge towards equilibrium. To address this point and the unrealistic fact that the Walrasian mechanism does not allow agents to trade until the equilibrium is reached, 
some economists have introduced the concept of \emph{non-t\^{a}tonnement} in which agents are allowed to trade before the equilibrium has been reached  \cite{fisher1989disequilibrium}. Whether such convergence dynamics, even 
in the presence of trading, should be identified to the dynamics of market price itself, is far from obvious. In fact, as mentioned in  \cite{mas1995microeconomic}, such a model should be thought of \emph{not as modelling the evolution of a supply-and-demand driven economy, but rather as a tentative trial-and-error process taking place in fictional time [...] [to find] the equilibrium level of prices}. 
We are thus speaking here of a \emph{transient} dynamics in an otherwise stable world. 

\item A second way of introducing dynamics is to consider a multiple period economy in which the supply and demand may evolve at each period, resulting in a new price (see for example \cite{mankiw2014principles}, Ch. 14). 
This collection of static equilibria is a rather weak notion of dynamics, that is closer to a \emph{quasi-static} evolution without transactions, in which the price is always the outcome of an equilibrium supply and an equilibrium demand. \cite{hicks1946value} noted that \emph{we shall find [...] that there is a way of reducing the dynamic problem into terms where it becomes formally identical with that of statics}, showing that something important is somehow missing. As we shall indeed argue below, this figment misses the essential point that when transactions occur, supply and demand curves are both immediately depleted, thus affecting subsequent transactions. Only when the time between market clearing auctions is large enough can the supply and demand curve again be considered in an equilibrium state prior to any further transaction. 

\item Following up on the last remark, our view is that a complete dynamic description must account for the evolution of supply and demand in an ever changing world, both in-between auctions/transactions and 
right when the auction takes place. This appears as a necessity if one wants to understand the formation of real prices, in particular in financial markets where supply and demand permanently interact, and where 
transactions prevent the supply and demand curves from being in an equilibrium at any point in time. To our knowledge, whereas many papers have worked in the direction of understanding price formation under continuous double auction \cite{gjerstad1998price,biais1993price,cason1996price,easley1993theories}, such a general dynamical theory of supply and demand is not available at this stage. This is what the present paper aims at achieving.
\end{enumerate}

As we shall see below, our proposal is at odds with classical approaches, which usually consider ``rigid" supply and demand curves, that shift uniformly  with respect to each other (see e.g. \cite{mankiw2014principles}, Ch.  14). In contrast, we propose a partial derivative equation that describes the evolution and deformation of the full price dependent supply and demand curves as the core ingredient of our model. 

\subsection{Financial economics: The Kyle model}\label{sec:kyle}

One strategy to model financial markets that has been popular since Kyle's seminal paper \cite{kyle1985continuous} is to consider markets as one- or several-period(s) equilibrium(a) between two or more (representative) agents, each representing a well-identified trading behaviour. In Kyle's original paper, an informed trader competes with a market maker who provides liquidity for every trade, in the presence of a noise trader who trades at random. This 
modelling strategy has been followed in many subsequent papers, where a small number of rational agents optimize a given utility function and maintain an ecological equilibrium. 

Whereas the results of such models often yield useful qualitative intuition, they miss -- in our opinion -- two essential features of markets. First, trading occurs in continuous time and reasoning in terms of periods 
(e.g. one period per trading day) is not appropriate in that respect. Second, the number of (representative) agents is usually pre-defined and small; typically, three strategies in Kyle's model. This has to be contrasted 
with real markets where agents are strongly heterogeneous, and the very idea of representative agents dubious. As a result, these equilibrium models do not reproduce some essential market features such as the square-root impact law and therefore probably miss some fundamental aspects of price formation. 

At variance with these usual models, the present paper suggests that relevant agent-based modelling should incorporate three essential features: (i) a one-dimensional definition of the price dynamics \textit{via} the order book, (ii) continuous time and (iii) heterogeneous agents. Based on these three ingredients, we define and solve below a particular tractable class of models that appears to capture faithfully some essential feature of price formation in continuous double-auction markets. But we believe that our results would hold for a much wider class of models based on the same ingredients.

\subsection{Financial economics: Models of the limit order book}\label{sec:fin}

Instead of the above ``macroscopic" considerations on the (unobservable) supply and demand, a whole branch of financial mathematics (concerned with ``market microstructure'') has recently emerged. 
The focus is on the actual evolution of the limit order book and of price formation on financial markets. The limit order book is described as a queueing system, in which buyers (resp. sellers) post quantities 
on a discrete price grid (the elementary price change is called the ``tick'') and wait for being executed. The buyers (resp. sellers) that offer the best price are then executed by aggressive sellers (resp. buyers) according to a \emph{first in, first out} policy. 
When the whole volume present on the best ask/bid queue is executed, then either some sell/buy volume replenishes the queue and the price is unchanged, or the queue is replenished by opposite side traders 
and the price moves by one tick. An obvious motivation for such research is to give practical answers to many questions from the financial industry (concerning optimal market making, optimal execution, optimal trading, etc.
 \cite{cont2013optimal,cartea2014optimal}) as well as from regulators (tick size, market ecology, market design and stability etc. \cite{robert2011new,laruelle2011optimal}). Much effort has been devoted to understand and model the mechanics of the limit order books, how it is affected by market design and the ecology of traders (in particular High Frequency Trading), and how it relates to macroscopic variables such as price volatility, etc \cite{foucault2013market,handa2003quote,hasbrouck2006empirical}. 

The availability of detailed data where all market events are recorded (i.e. trades, quotes, cancellations, etc.) has generated a flurry of empirical papers, describing many aspects of price formation at the microstructural level 
for a review see \cite{biais1995empirical,bouchaud2008markets}. Correspondingly, a host of stylized models of the order book have appeared, with different starting points and objectives.  
For example, ``zero-intelligence'' models  \cite{smith2003statistical, bouchaud2002statistical,farmer2005predictive,cont2010stochastic,gareche2013fokker} form an important class of models of the order book, where one assumed that agents act mechanically (rather than strategically) leading to
simple Poissonian statistics for the order flow. Although obviously too simple to account for what goes on in financial markets, such models reveal some interesting relationships between observables (spreads, volatility, activity, etc.) \cite{farmer2005predictive,cont2013price}. Much more elaborate models have also been developed, taking into account the heterogeneity, strategies and preferences of market participants \cite{foucault1999order, rocsu2009dynamic, rosu2014liquidity, maglaras2011multiclass,lachapelle2013efficiency}, some including the queues behind the best buy/sell prices \cite{huang2014simulating}. 

The present paper is clearly partly inspired by the above strand of papers on real limit order books, in particular \cite{smith2003statistical,bouchaud2002statistical}. However, we depart from these models on one very fundamental issue. 
Instead of trying to describe the evolution of the \emph{visible} order book (where only a tiny fraction of the outstanding liquidity is revealed, and whose dynamics is dominated by highly strategic market-makers/HFT), we want to describe the much deeper and much slower ``latent'' order book, introduced in  \cite{Toth:2011}, that contains all buy/sell {\it intentions}, whether displayed or not by market participants. In other words, we model the true underlying supply and demand curves that would materialize if the transaction price was to move closer to the reservation prices. The distinction between the visible limit order book 
(which, as stated above, gives a very poor indication on liquidity at larger scales) and the true supply and demand curves is absolutely crucial for all that follows. The model described below is 
a generalisation of the ideas introduced in \cite{Toth:2011,Iacopo:2013} and in \cite{donier2014fully}. It builds upon the intuition that  
agents can revise their reservation prices in an heterogeneous manner, introduced long ago in \cite{bak1997price} and recently revisited in completely different contexts in \cite{lasry2007mean, lehalle2011high} and in \cite{Toth:2011}. The motivation of the latter paper
was to explain the universal concave (``square-root'') impact of directional trade sequences mentioned in the introduction, that deeply challenges standard equilibrium models. 

In summary, the aim of the present paper is to reconcile the insights gained by the financial literature on price formation with a more Walrasian view of supply and demand that provides us with a macroscopic theory 
of price formation. We believe that this reconciliation has important conceptual consequences from an economic perspective (in particular in emphasizing the {\it dynamical aspects} of price formation and liquidity), as well as practical implications for market design and regulation (in particular concerning the crucial issue of market stability).

\section{A dynamic theory of the supply \& demand curves}\label{sec:dynamics}

\subsection{Definitions}

The classical supply and demand curves $S(p,t)$ and $D(p,t)$ (SD) represent respectively the amount of supply and demand that would reveal themselves 
if the price were to be set to $p$ at time $t$. In classical Walrasian auctions, the equilibrium price $p_t^*$ is then set to the value that matches both quantities so that $D(p_t^*,t)=S(p_t^*,t)$. This equilibrium is unique provided the curves are strictly monotonous\footnote{By definition, or simply by common sense, the demand curve is a decreasing function of the price whereas the supply curve is increasing.}. The supply and demand curves, as well as the resulting equilibrium price, are represented on Fig. \ref{fig:SD} (left).

In order to define the dynamics of the supply and demand curves, we also introduce the \emph{marginal supply and demand curves} (MSD), on which we will focus in the 
rest of this paper. They are defined as the derivative of the SD curves 

\be
\begin{aligned}
&\rho_S(p,t)=\partial_p S(p,t)\geq 0;\\ 
&\rho_D(p,t)=-\partial_p D(p,t) \geq 0,
\end{aligned}
\ee
\noindent with the following interpretation: For any price $p$, $\rho_S(p,t)\text{d}p$ (resp. $\rho_D(p,t) \text{d}p$) is, at time $t$, the quantity of supply (resp. demand) that would materialize if the price changed from $p$ to $p+\text{d}p$ (resp. $p-\text{d}p$). The MSD curves can thus be seen as the density of supply and demand intentions in the vicinity of a given price. Fig. \ref{fig:SD} (right) shows MSD curves corresponding to the SD curves: Higher MSD levels correspond to larger slopes for the SD curves.

In the Walrasian story, supply and demand pre-exist and the Walrasian auctioneer gropes ({\it t\^atonne}) to find the price $p_t^*$ that maximizes the amount of possible transactions. The auction then takes place at time $t$ and removes instantly all matched orders. Assuming that all the supply and demand intentions close 
to the transaction price were revealed before the auction and were matched, the state of the MSD just after the auction is simple to describe, see Figs. \ref{fig:SD} \&  \ref{fig:scheme}:

\be\label{eq:after}
\left\lbrace 
\begin{array}{ccc}
\rho_S(p,t^+) &= &\rho_S(p,t^-) \quad (p > p_t^*)\\
 & = &0  \quad (p \leq p_t^*)\\
\rho_D(p,t^+) &= &\rho_D(p,t^-) \quad (p < p_t^*)\\
&=&0  \quad (p \geq p_t^*).\\
\end{array}
\right.
\ee

But what happens next, once the auction has been settled? So far the story does not tell (to the best of our knowledge). The aim of the following is to set up a 
general framework for the dynamics of the supply and demand curves. This will allow us to describe, among other questions, 
how the supply and demand curves evolve from the truncated shape given by Eq. (\ref{eq:after}) up to the next auction  
at time $t + \tau$ (where $\tau$ is the inter-auction time).

\subsection{General hypotheses about the behaviour of agents}\label{sec:behaviour}

The theory that we present here relies on weak and general assumptions on agents behaviours that translate into a simple and universal 
evolution of the MSD curves, with only very few parameters\footnote{In fact, as shown in  \cite{donier2014fully} and below, only two parameters suffice to describe the problem in the vicinity of the price: 
one is the price volatility, and the other one is related to market activity (traded volume per unit time).}. The MSD curves aggregate the intentions of all agents, which would materialize in the 
``real'' order book if it was not for fear of being picked off by more informed traders, or of revealing some information to the market. This is why the MSD
curves were called the ``latent'' order book (LOB) in Refs. \cite{Iacopo:2013,Iacopo:2014,donier2014fully}, as initially proposed in  \cite{Toth:2011}. 

We will assume that there is a so-called ``fundamental'' price process $\widehat p_t$ which is only partially known to agents, in a sense clarified below (see Sec. \ref{sec:discussion}). For simplicity, we will also posit that $\widehat p_t$ is an additive Brownian motion. In the absence of transactions, the MSD curves evolve according to three distinct mechanisms, that we model as follows:

\begin{itemize}

\item New intentions, not present in the supply and demand before time $t$, can appear. The probability for new buy/sell intentions to appear between $t$ and $t + {\rm d}t$ and between prices $p$ and $p+{\rm d}p$ 
is chosen to be $\omega_\pm(p - \widehat p_t)$, where $\omega_+(x)$ is 
a decreasing function of $x$ and $\omega_-(x)$ is an increasing function of $x$.

\item Already existing intentions to buy/sell at price $p$ can simply be cancelled and disappear from the supply and demand curves. The probability for an existing buy/sell intention 
around price $p$ to disappear between $t$ and $t + {\rm d}t$ is chosen to be $\nu_\pm(p - \widehat p_t)$.

\item Already existing intentions to buy/sell at price $p$ can be revised. Between $t$ and $t + {\rm d} t$, each agent $i$ revises his/her reservation price $p^i$ to $p^i + \beta^i {\rm d}\xi_t + {\rm d}W_{i,t}$, where ${\rm d} \xi_t$ 
is common to all $i$, representing some public information. $\beta^i$ is the sensitivity of agent $i$ to the news, which we imagine to be a random variable 
from agent to agent, with a mean normalized to $1$. Some agents may over-react ($\beta^i > 1$), others under-react ($\beta^i < 1$). 
The idiosyncratic contribution ${\rm d} W_{i,t}$ is an independent Wiener noise both across different agents and in time, 
with distribution of mean zero\footnote{One could generalize the calculations below to the case where the mean is non zero (modelling for example the tendency
of agents to revise their reservation price in the direction of the traded price). This would affect none of the conclusions below, at least in the limit where the 
inter-auction time $\tau$ becomes very small.} and variance $\Sigma_i^2 {\rm d} t$, that may depend on the agent (some agents might be more ``noisy'' than others). 

\end{itemize}

We will furthermore assume that the ``news'' term ${\rm d}\xi_t$ is a Wiener noise of variance $\sigma^2 \d t$, corresponding to a Brownian motion for the fundamental price 
$\widehat p_t = \int^t  {\rm d}\xi_{t'}$ with volatility $\sigma$. 
Normalising the mean of the $\beta^i$'s to unity thus corresponds to the assumption that agents are on average unbiased in their interpretation of the news -- i.e. their intentions remain
centred around the fundamental price $\widehat p_t$ in the course of time -- but see the expanded discussion of this point in Sec. \ref{sec:discussion}. 

Our central assumptions are \emph{heterogeneity}, together with the hypothesis that idiosyncratic behaviours ``average out'' in the limit of a very large number of participants, i.e., no single agent accounts for a finite fraction 
of the total supply or demand.  While not strictly necessary, this assumption leads to a deterministic aggregate behaviour and allows one to gloss over some rather involved mathematics.

\subsection{The model in terms of optimizing agents}
\label{sec:agent-opt}

The above assumptions might appear obscure to those used to think in terms of rational optimizing agents and equilibria. Here we rephrase these assumptions in a language closer to standard economic intuition.

We consider an \emph{open} economic system, in which many heterogeneous, infinitesimal agents operate. Each agent $i$ has a certain utility 
$\mathcal{U}_i(p,\theta|\widehat{p}_t^i,{\cal F}_t)$ for buying ($\theta=+1$) or selling ($\theta=-1$) a unit (small) quantity 
at price $p$, given his/her estimate of the fundamental price $\widehat{p}_t^i$ and all the information about the rest of the world, available at time $t$, encoded in 
${\cal F}_t$. The third option available to agent $i$ is to be inactive ($\theta=0$), in which case the number of goods s/he owns remains constant. 
Agents are heterogeneous in the sense that both their utility function and their estimates of the fundamental price are different; one can think of them as random members 
of some adequate statistical ensembles. For the sake of simplicity, we consider no interest rate and no risk of any kind.

At time $t$, each agent computes his optimal action $p_t^i, \theta_t^i$ as the result of the following optimisation program:
\be\label{opt}
(p_t^i,\theta_t^i) = \underset{p,\theta}{\text{argmin}}~~\mathcal{U}_i(p,\theta|\widehat{p}_t^i,{\cal F}_t).
\ee
Because of the random evolution of the outside world summarized by $\widehat{p}_t^i,{\cal F}_t$, the value of $\theta_t^i \in \{-1,0,+1\}$ can change between
$t$ and $t + \d t$. For the sake of simplicity, we assume that the change of the state of the world in time $\d t$ is never so large as to induce direct 
transitions from $\mp 1 \to \pm 1$ without pausing at $0$. Hence, between $t$ and $t + \d t$, the following transitions (or absence thereof) are possible:
\begin{itemize}
\item $0 \to 0$: this clearly induces no change in the MSD curves;
\item $0 \to \pm 1$: in this case, agent $i$ previously absent from the market becomes either a buyer or a seller, with reservation price $p_t^i$ given by
Eq. (\ref{opt}). The assumption that agents are heterogeneous translates in a model where this event is a Poisson process with some arrival rate $\omega_\pm(p)$;
\item $\pm 1 \to 0$: in this case, agent $i$ previously present in the market as a buyer or a seller, decides to become neutral, which is modelled as a Poisson process with some cancellation rate $\nu_\pm(p)$;
\item $\pm 1 \to \pm 1$: in this case, a buyer/seller remains a buyer/seller, but may change his/her reservation price because the solution of Eq. (\ref{opt})
has changed. Writing $p_t^i = f_i(\widehat{p}_t^i,t)$, where $f$ is a regular function if $\mathcal{U}_i$ is regular enough, and applying It\^{o}'s lemma, one finds:
\be
\begin{aligned}
\d p_t^i &= \frac{\partial f_i}{\partial t} \d t + \frac{\partial f_i}{\partial p} \d \widehat{p}_t^i + \frac{\sigma_i^2}{2}
\frac{\partial^2 f_i}{\partial p^2} \d t \\
&= \alpha_t^i \d \widehat{p}_t^i + \gamma_t^i \d t.
\end{aligned}
\ee
The drift term $\gamma_t^i$ will play little role in the following (see previous footnote), and we neglect it henceforth. In order to recover the specification
of the above section, we further decompose the price revision $\d p_t^i = \alpha_t^i \d \widehat{p}_t^i$ into a \emph{common} component $\beta^i {\rm d}\xi_t$ 
and an \emph{idiosyncratic} component ${\rm d}W_{i,t}$ as above.
\end{itemize}

Therefore, the mechanism proposed in the above section indeed describes the behaviour of an open system of \emph{infinitesimal} and \emph{heterogeneous} market
participants. Note that we do not need to distinguish between fundamental investors, noise traders and market makers, as for example in the Kyle model \cite{kyle1985continuous}. This is due to our assumption that the market contains a very large number of participants, in which case the MSD curves are 
continuous. Discretization effects (in price and in quantity) would open gaps in the MSD curves, and specific market makers would then be needed to ensure continuous, orderly trading. Finally, and quite importantly, the price dynamics in the above setting is arbitrage free (see \cite{donier2014fully}). There is therefore 
no optimal strategic component that is missing from the above utility maximisation program.

\subsection{The ``free evolution'' equation for the MSD curves}\label{sec:free}

Endowed with the above hypothesis, one can derive stochastic partial differential equations for the evolution of the marginal supply ($\rho_S(p,t)=\partial_p S(p,t)$) and the marginal 
demand ($\rho_D(p,t)=-\partial_p D(p,t)$) in the absence of transactions  \cite{donier2014fully}.
It turns out that, as expected, these equations take a simpler form in the reference frame of the (moving) fundamental price $\widehat p_t$. Introducing the shifted price $y = p - \widehat p_t$, 
one finds \cite{donier2014fully}:\footnote{See also \cite{lasry2007mean, lehalle2011high} for similar ideas in the context of mean-field games. 
Note that Eq. (\ref{eq:dynamics}) is strictly valid when $\rho_S(p,t)$ and $\rho_D(p,t)$ are be interpreted as the marginal supply and demand curves averaged over the noise processes. 
Otherwise some noisy component remains, see e.g.  \cite{dean1996langevin}.}
 \begin{eqnarray}
\label{eq:dynamics}\nonumber
\partial_t \rho_D(y,t) &=&  {\cal D} \partial^2_{yy} \rho_D(y,t)  - \nu_+(y) \rho_D(y,t) + \omega_+(y); \\
\partial_t \rho_S(y,t) &=& \underbrace{\D \partial^2_{yy} \rho_S(y,t)}_{\text{Updates}} - 
\underbrace{\nu_-(y) \rho_S(y,t)}_{\text{Cancellations}} + \underbrace{\omega_-(y)}_{\text{New orders}},
\end{eqnarray}
where ${\cal D} = \frac12 [\E_i(\Sigma_i^2) + \sigma^2 {\text{Var}}(\beta^i)]$, i.e. part of the diffusion term comes from the purely idiosyncratic ``noisy" updates of agents ($\E_i(\Sigma_i^2)$),  and another part comes from the 
inhomogeneity of their reaction to news ($\sigma^2 {\text{Var}}(\beta^i))$, which indeed vanishes if all $\beta^i$'s are equal to unity.\footnote{Here we neglect the possibility that buyers and sellers update their
price differently, but one could make a distinction between a ${\cal D}_+$ and a ${\cal D}_-$, or even make ${\cal D}$ price/time dependent.} 

These equations, that are at the core of the present paper, describe the structural 
evolution of supply and demand around the fundamental price $\widehat{p}_t$. Notice however that $\widehat{p}_t$ has disappeared from the above equations. The dynamics of the MSD curves can be treated independently from the dynamics of the price itself, provided one describes the MSD in the reference frame of the price. There is however a direct relationship between the price volatility $\sigma$ and the diffusion coefficient ${\cal D}$, as expressed above and noted in  \cite{donier2014fully}.

Interestingly, whereas the price is random and follows a rough path (typically a Brownian motion), the structural part is deterministic and smooth, thanks to the assumption of ``infinitesimal'' orders (that can be made rigourous by considering an appropriate scaling for system parameters that corresponds to a hydrodynamic limit, see \cite{gao2014hydrodynamic}). 

The above equations for $\rho_D(y,t)$ and $\rho_S(y,t)$ are linear and can be formally solved in the general case, starting from an arbitrary initial condition such as Eq. (\ref{eq:after}), using a spectral decomposition of
the evolution operator. This general solution is however not very illuminating, and we rather focus here in the special case where $\nu_\pm(y) \equiv \nu$ does not depend on $y$ nor on the side of the latent order book. The general solution can then be written in a fairly transparent way, as:
\be \label{eq:gen-sol}
\rho_{S,D}(y,t) = \int_{-\infty}^{+\infty} \frac{\d y'}{\sqrt{4 \pi \D t}} \, \rho_{S,D}(y',t=0^+) e^{-\frac{(y'-y)^2}{4 \D t} - \nu t} 
+ \int_0^t \d t' \int_{-\infty}^{+\infty} \frac{\d y'}{\sqrt{4 \pi \D (t-t')}} \, \omega_{\pm}(y') e^{-\frac{(y'-y)^2}{4 \D (t-t')} - \nu (t-t')},
\ee
where $\rho_{S,D}(y,t=0^+)$ is the initial condition, i.e. just after the last auction. 

We will now explore the properties of the above solution at time $t= \tau^-$, i.e., just before the next auction, in the two asymptotic limits $\tau \to \infty$, corresponding to very infrequent auctions, and $\tau \to 0$, corresponding to continuous time auctions.

\section{Discrete Auctions and Price Impact}\label{sec:theory}

The aim of this section is to show that the shape of the marginal supply and demand curves can be fully characterized in the limit of very infrequent auctions (corresponding to Walras' auctions) and in the opposite limit of nearly continuous time
auctions (corresponding to financial markets),  and describe the transition between the two limits. The upshot is that while the liquidity around the auction price is in general finite and leads 
to a linear impact using the standard argument in Eq. (\ref{Kyle-lambda1}) above, this  liquidity vanishes as $\sqrt{\tau}$ when the inter-auction time $\tau \to 0$. This signals the breakdown of linear impact and, as shown at the end of the section, 
its replacement by the square-root law mentioned in the introduction.

\subsection{Walras, or the limit of infrequent auctions}

Letting $t=\tau \to \infty$ in the above Eq. (\ref{eq:gen-sol}), one immediately sees that the first term disappears, meaning that one reaches a {\it stationary solution} $\rho^{\text{st.}}_{S,D}(y)$ that is independent
of the initial condition. The second term can be simplified further to give the following general solution:
\be\label{eq:st-sol}
\rho^{\text{st.}}_{S,D}(y) =  \frac{1}{2 \sqrt{\nu \D}} \int_{-\infty}^{+\infty} \d y' \omega_{\pm}(y') \, e^{-\sqrt{\frac{\nu}{\D}}|y'-y|}.
\ee
A particularly simple case is when $\omega_{\pm}(y)=\Omega_\pm e^{\mp \mu y}$, meaning that buyers(/sellers) have an exponentially small probability to be interested in a transaction at high/low prices. In this toy-example, one
readily finds that a stationary state only exists when $\nu > \D \mu^2$ and reads:
\be
\rho^{\text{st.}}_{S,D}(y) = \frac{\Omega_\pm}{\nu - \D \mu^2} e^{\mp \mu y}.
\ee
Other forms for $\omega_{\pm}(y)$ can be investigated as well, for example $\omega_{\pm}(y) = \omega_{\pm}^0\mathbbm{1}_{\{y<>0\}}$ which yields:
\be
\rho^{\text{st.}}_{S,D}(y) = \frac{\omega_{\pm}^0}{2\nu}\left[ 1\pm\text{sign}(y)(1-e^{-\sqrt{\nu/D}\mid y \mid}) \right],
\ee
that we will use in Figs. \ref{fig:books} and \ref{fig:vanishing}. The shape of $\rho^{\text{st.}}_{S,D}(y)$ is generically the one shown in Fig. \ref{fig:SD} with an overlapping region where buy/sell orders coexist. 
The auction price $p^*_\tau = \widehat p_\tau + y^*$ is determined by the condition $D(p_\tau^*,\tau^-)=S(p_\tau^*,\tau^-)$, or else:
\be
\int_{y^*}^\infty \d y \rho^{\text{st.}}_{D}(y) = \int_{-\infty}^{y^*} \d y \rho^{\text{st.}}_{S}(y) \equiv  v^*,
\ee
where $v^*$ is the volume exchanged during the auction. For the simple exponential case above, this equation can be readily solved as:  
\be \label{eq:lin-imp}
y^* = \frac{1}{2\mu} \ln \frac{\Omega_+}{\Omega_-},
\ee
with a clear interpretation: if the new buy order intentions accumulated since the last auction happen to outsize the new sell intentions during the same period, the auction price will exceed the fundamental price, 
and vice-versa. This pricing error is expected to be small if the order book is observable during the inter-auction period, since in that case $\Omega_+$ and $\Omega_-$ will track each other and remain close. 
Otherwise, one expects the imbalance to invert in the next period, leading to a kind of ``bid-ask bounce'' well known in the context of market microstructure. One can also compute the volume exchanged 
during the auction $v^*$. One finds: 
\be
v^* = \frac{\sqrt{\Omega_+ \Omega_-}}{\mu(\nu - \D \mu^2)}.
\ee
Just after the auction, the MSD curves start again 
from $\rho^{\text{st.}}_{S,D}(y)$, truncated below (resp. above) $y^*$, as in Eq. (\ref{eq:after}).

Let us now turn to price impact in this model. From Eq. (\ref{eq:st-sol}), it is immediate that for any clearing price $y^*$, both $\rho^{\text{st.}}_{S}(y^*)$ and $\rho^{\text{st.}}_{D}(y^*)$ are strictly positive. 
This would remain true even if the dependence on $y$ of cancellation rate $\nu_\pm(y)$ was reinstalled. 
The general argument given in the introduction therefore predicts a {\it linear impact} for an extra buy/sell quantity given by:
\be\label{Kyle-lambda2}
{\cal I}(Q) = \pm \lambda Q; \qquad  \qquad \lambda =\frac{1}{\rho^{\text{st.}}_{S}(y^*)+\rho^{\text{st.}}_{D}(y^*)}.
\ee
For the exponential case, this again takes a simple form:
\be
\lambda = \frac{{ \nu-\D \mu^2}}{2 \sqrt{\Omega_+ \Omega_-}},
\ee
whereas for a general symmetric order flow $\omega_{+}(y)=\omega_{-}(-y)$, $y^*$ is obviously equal to zero, leading to:
\be
\lambda = \frac{\sqrt{\nu \D}}{\int_{-\infty}^{+\infty} \d y' \omega(y') e^{-\sqrt{\frac{\nu}{\D}}|y'|}}.
\ee
For $\omega_{\pm}(y) = \omega^0\mathbbm{1}_{\{y<>0\}}$, one obtains the simple and intuitive result: 
\be
\lambda = \frac{\nu}{\omega_0},
\ee
i.e. that the market liquidity, measured by $\lambda^{-1}$, grows linearly with the rate of incoming orders and inversely proportionally to the 
cancellation rate.

The main point of the present section is that when the inter-auction time is large enough, each auction clears an equilibrium supply with an equilibrium demand, with very simple and predictable outcomes. This corresponds to the quasi-static dynamics discussed in item 2., Section ~\ref{sec:eco}, and to the standard representation of market dynamics in the Walrasian context, since in this case only the long-term properties of supply and demand matter and the whole transients are discarded. The next section will depart from this limiting case, by introducing a finite inter-auction time such that the transient dynamics of supply and demand becomes a central feature in the theory.

\subsection{High frequency auctions}

We will now investigate the alternative limit where the inter-auction time $\tau$ tends to zero. Since all the supply (resp. demand) curve left (resp. right) 
of the auction price is wiped out by the auction process, one expects intuitively that after a very small time $\tau$, the density of buy/sell orders in the
immediate vicinity of the transaction price will remain small. We will show that this is indeed the case, and specify exactly the shape of
the stationary MSD after many auctions have taken place. Consider again Eq. (\ref{eq:gen-sol}) just before the $n+1$th auction at time $(n+1) \tau^-$, in the
case where the flow of new orders is symmetric, i.e. $\omega_{+}(y)=\omega_{-}(y)$, such that the transaction price is always at the fundamental price ($y^* = 0$). We will focus on
the supply side and postulate that $\rho_S(y,t=n \tau^-)$ can be written, in the vicinity of $y=0$,  as \footnote{This approximation happens to be exact in the particular setting considered in  \cite{donier2014fully}.}:
\be\label{eq:iteration}
\rho_S(y,t=n \tau^-) = \sqrt{\tau} \, \phi_n\left(\frac{y}{\sqrt{\D \tau}}\right)+ O(\tau)
\ee
when $\tau \to 0$ (and symmetrically for the demand side). Plugging this ansatz into Eq. (\ref{eq:gen-sol}), making the change of variable $y' \to \sqrt{\D \tau} w$ 
and taking the limit $\tau \to 0$ leads to the following iteration equation, exact up to order $\sqrt{\tau}$:\footnote{An extra correction of order $\sqrt{\tau}$ would appear if a drift term was added to Eqs. (\ref{eq:dynamics}).}
\be
\phi_{n+1}(u) = \int_0^{+\infty} \frac{\d w}{\sqrt{4 \pi}} \phi_n(w) e^{-(u-w)^2/4} + \sqrt{\tau} \, \omega(0) + O(\tau).
\ee
Note that $\nu$ has entirely disappeared from the equation (but will appear in the boundary condition, see below), and only the value of $\omega$ close to the transaction price is relevant at this order.

After a very large number of auctions, one therefore finds that the stationary shape of the demand curve close to the price and in the limit $\tau \to 0$ is given by the non-trivial solution of the following 
fixed point equation:
\be\label{eq:atkinson}
\phi_\infty(u) = \int_0^{+\infty} \frac{\d w}{\sqrt{4 \pi}} \phi_\infty(w) e^{-(u-w)^2/4},
\ee
supplemented by the boundary condition $\phi_\infty(u \gg 1) \approx \L \sqrt{\D} u$, where $\L$ is a constant to be determined below. [Note that the solution of Eq. (\ref{eq:atkinson}) is determined up to a 
multiplicative factor that must be fixed by some external condition].
 
 \begin{figure}[t!]
    \begin{center}
        \includegraphics[height=7.5cm]{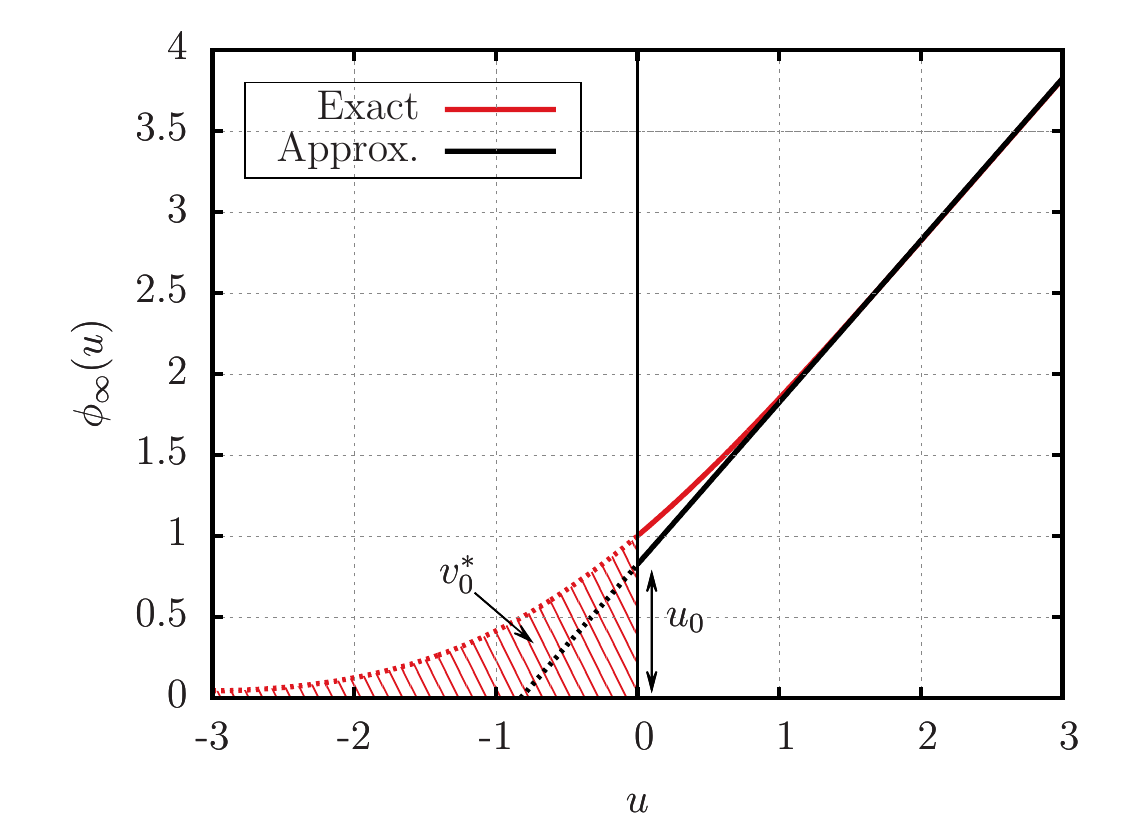}
    \end{center}
    \caption{Graph of the normalized exact solution $\phi_\infty(u)$, and its affine approximation. The whole picture must be rescaled by a factor $\sqrt{\tau}$ to recover the order book when the inter-auction time is $\tau$. The hatched region corresponds to the volume to be executed, and therefore scales with $\tau$.}
    \label{fig:atkinson}
\end{figure}

 \begin{figure}[t!]
    \begin{center}
        \includegraphics[height=7.5cm]{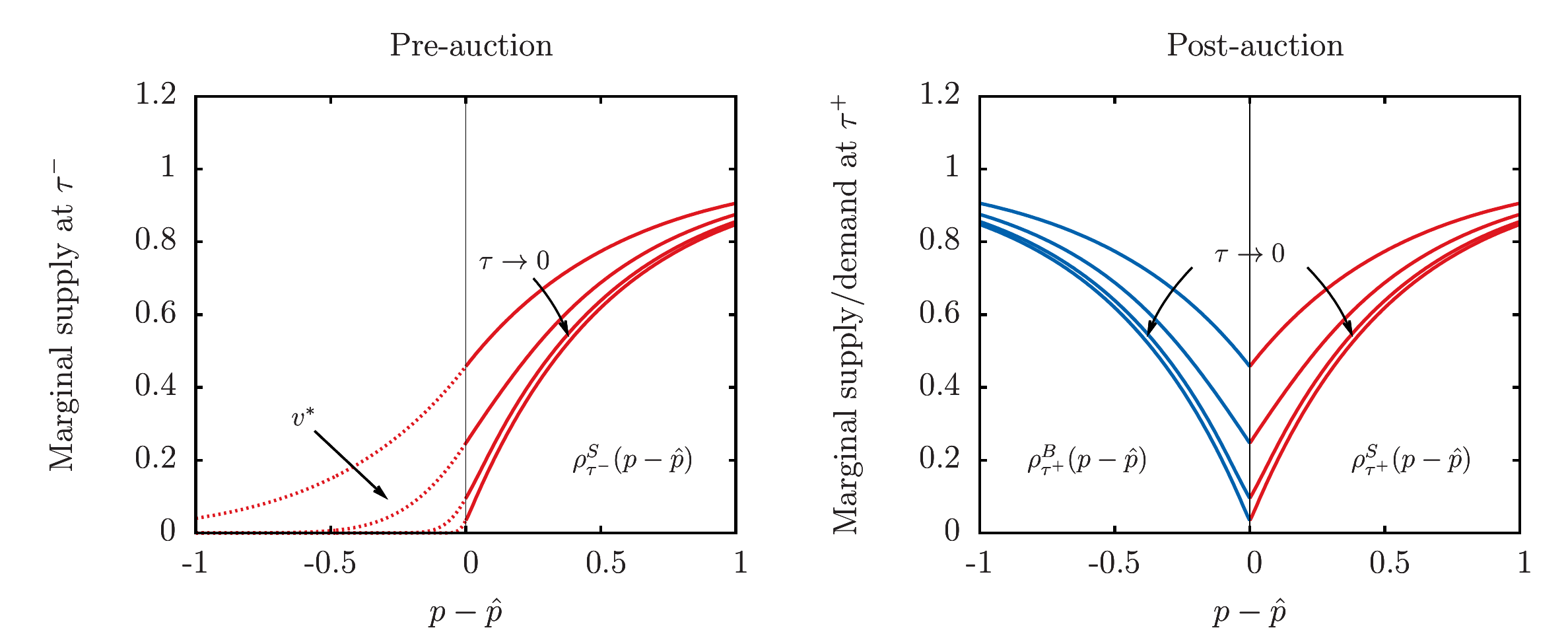}
    \end{center}
    \caption{Left: Shape of the marginal supply curve immediately before the auctions, for different inter-auction times $\tau$, in the case $\omega_{\pm}(y) = \omega_{\pm}^0\mathbbm{1}_{\{y<>0\}}$. Right: Shape of the MSD immediately after the auctions, again for different 
    inter-auction times $\tau$. Note that as $\tau \to 0$, the MSD acquires a characteristic V-shape.}
    \label{fig:books}
\end{figure}

Equation (\ref{eq:atkinson}) is of the Wiener-Hopf type and its analytical solution can be found in  \cite{atkinson1974wiener,boersma1974note}. We plot numerically this solution in Fig. \ref{fig:atkinson}; it is seen to be 
numerically very close to an affine function for $u > 0$: $\phi_\infty(u) \approx \L \sqrt{\D} (u + u_0)$ with $u_0 \approx 0.824$. In summary, the stationary shape $\rho_{S,\text{st.}}(y)$ of the marginal supply curve in the frequent auction limit $\tau \to 0$ and close to the transaction price ($y = O(\sqrt{\D \tau})$, 
has a {\it universal shape}, independent of the detailed specification of the model (i.e., the functions $\nu_\pm(y)$ and $\omega_\pm(y)$). This supply curve is given by 
$\sqrt{\tau} \phi_\infty(y/\sqrt{\D \tau})$, which can itself be approximated by a simple affine function that will fully suffice for the purpose of the present paper:
\be\label{eq:final}
\rho_{S,\text{st.}}(y \geq 0) \approx  \L (y + y_0); \qquad y_0 = u_0 \sqrt{\D \tau}; \qquad (\tau \to 0),
\ee
and similarly for $\rho_{D,\text{st.}}(y)$, see Fig. \ref{fig:books}. The detailed interpretation of this result -- in terms of market liquidity and price impact -- will be given below. 

We however still need to find the value of $\L$. This is done by comparing with the stationary solution $\varphi_{\text{st.}}(y)$ of Eq. (\ref{eq:dynamics}) that satisfies the boundary solution 
$\varphi_{\text{st.}}(0)=0$ (valid for $\tau=0$). For $\nu_\pm(y) = \nu$, $\varphi_{\text{st.}}(y)$ can be computed explicitly and is given by:
\be
\varphi_{\text{st.}}(y) =  \frac{1}{\D}  e^{-\sqrt{\nu/\D}\, y} \int_0^y {\rm d}y' e^{2\sqrt{\nu/\D} \,y'} \int_{y'}^\infty {\rm d}y'' e^{-\sqrt{\nu/\D} \,y''}\omega(y'').
\ee
Expanding $\varphi_{\text{st.}}(y)$ for small $y$ (but still much larger than $\sqrt{\D \tau}$) finally leads to:
\be
\varphi_{\text{st.}}(y) \approx \L y; \qquad \L = \frac{1}{\D} \int_{0}^\infty {\rm d}y' e^{-\sqrt{\nu/\D} \,y'}\omega(y'),
\ee
where $\L$ can be seen as a measure of the market liquidity (see  \cite{donier2014fully} and below).  Again in the simple case $\omega_{\pm}(y) = \omega^0\mathbbm{1}_{\{y<>0\}}$, one
finds:
\be
\L = \frac{\omega_0}{\sqrt{\nu {\cal D}}}.
\ee
Therefore, liquidity increases with the order arrival rate and decreases with their cancellation rate, as above, but also decreases 
with the diffusion constant ${\cal D}$ that can be loosely identified with market volatility (see the discussion above).

 \begin{figure}[t!]
    \begin{center}
        \includegraphics[height=7cm]{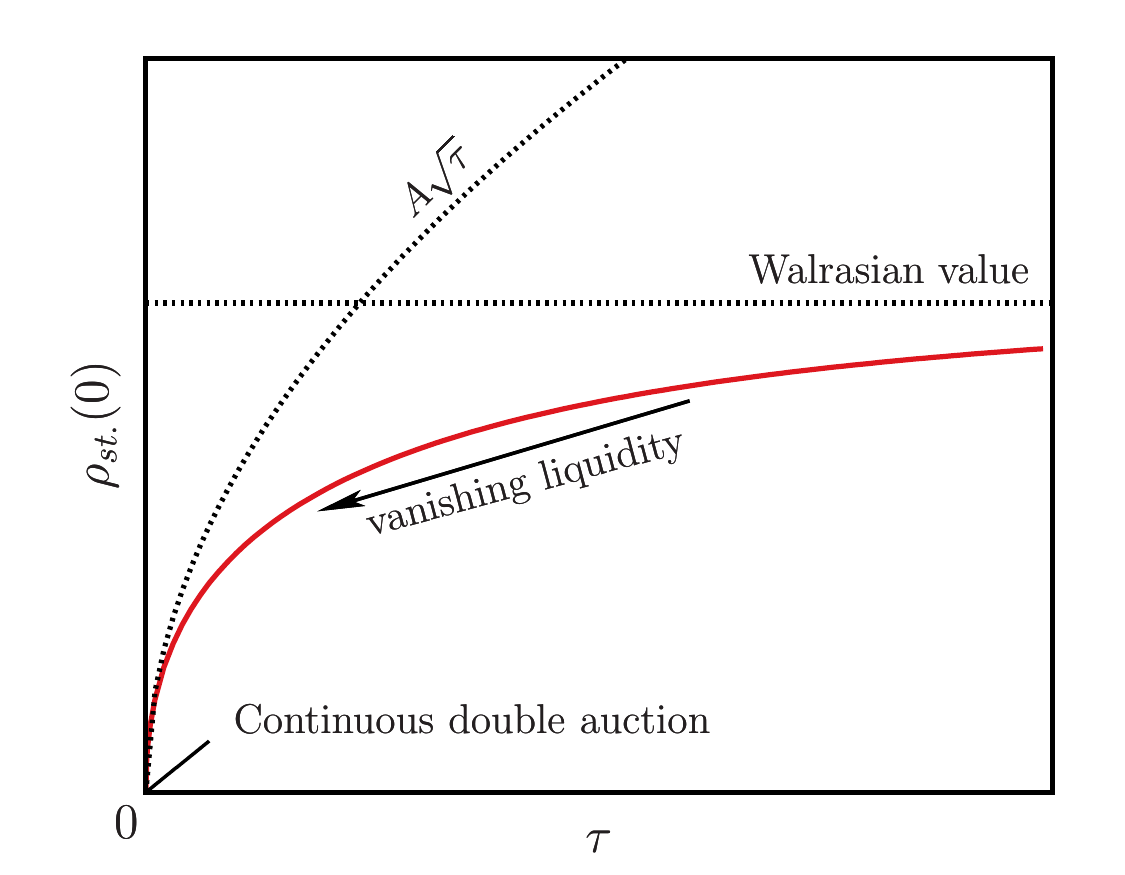}\includegraphics[height=7cm]{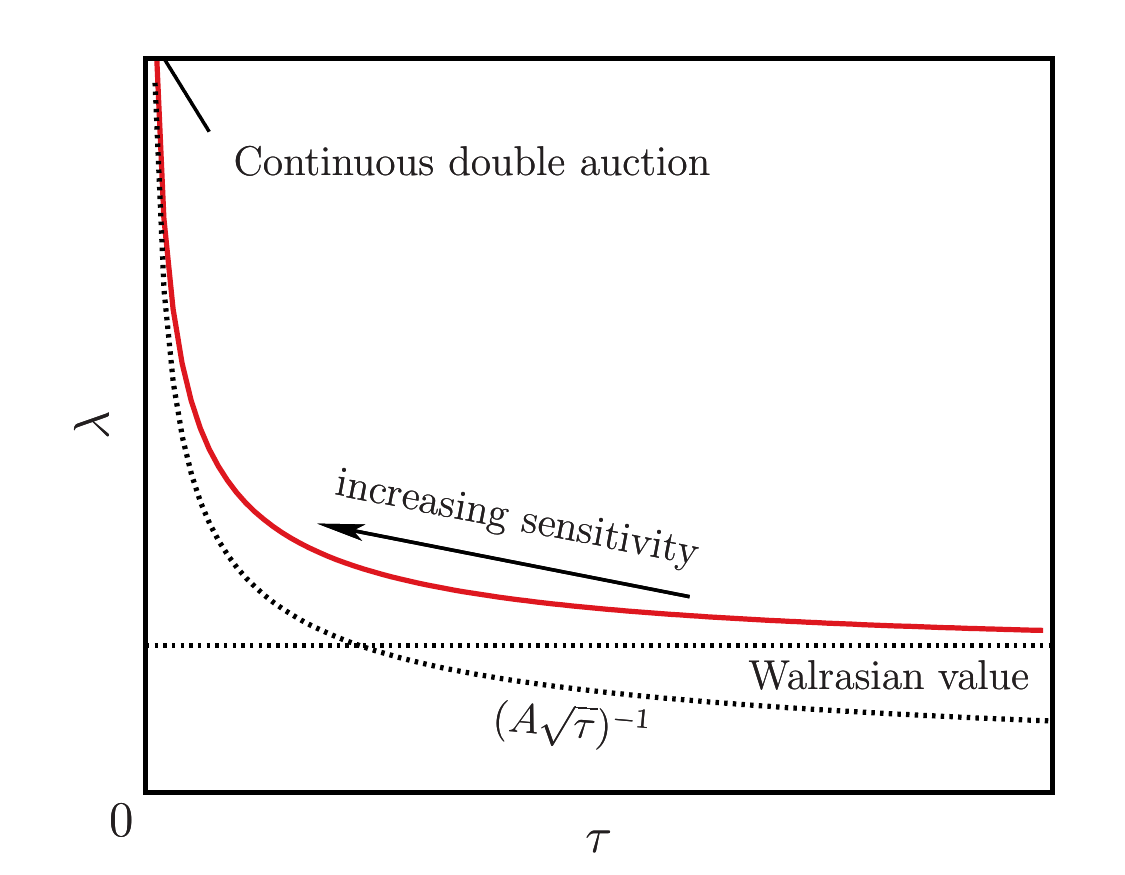}
    \end{center}
    \caption{Left: As the inter-auction time $\tau$ decreases, the liquidity close to the price, $\rho_{\text{st.}}(y = 0)$ decreases from its finite Walrasian value to zero in the case of continuous double auctions. Right: Conversely, the relative impact of (small) additional volumes $\lambda$ diverges for continuous double auctions ($\lambda \to \infty$ when $\tau \to 0$), eventually leading to the square-root impact law. }
    \label{fig:vanishing}
\end{figure}

Coming back to Eq. (\ref{eq:final}), one notes that Kyle's $\lambda$ behaves as $\lambda^{-1}\equiv 2\rho_{S,\text{st.}}(y=0) \propto \sqrt{\tau}$, which is the pivotal result of the present paper. 
It means that the marginal supply and demand at the transaction price becomes very small around the transaction price as the auction frequency increases. 
Intuitively, this is due to the fact that close to the transaction price, liquidity has no time to rebuild
between two auctions. From the point of view of impact, the divergence of Kyle's $\lambda$ as $1/\sqrt{\tau}$ means that the 
auction price becomes more and more susceptible to any imbalance between supply and demand. We show in Fig. ~\ref{fig:vanishing} (left) $\rho_{\text{st.}}(y=0)$ in the special case where 
$\omega_{\pm}(y) = \omega^0\mathbbm{1}_{\{y<>0\}}$, illustrating how liquidity vanishes as $\tau \to 0$ as well as (right) the corresponding impact parameter $\lambda(\tau)$ that diverges in this limit. 
One can thus see how, by increasing the auction frequency, one smoothly departs from the Walrasian equilibrium scenario to reach the limit of continuous double auction corresponding to modern financial markets.

The last item we need is the shape of the supply curve {\it below} the transaction price just before the next auction, that gives the amount of supply/demand on the ``wrong'' side of the price, i.e. precisely 
the volume exchanged at the auction. Using the simple affine approximation Eq. (\ref{eq:final}), one finds:
\be
\rho_{S,\text{st.}}(y < 0) \approx  \L \int_0^{+\infty} \frac{\d y'}{\sqrt{4 \pi \D \tau}} \, (y' + y_0) e^{-\frac{(y'-y)^2}{4 \D \tau}},  
\ee
or, again setting $y = -u \sqrt{\D \tau}$ and $y' = w\sqrt{\D \tau}$,
\be\label{eq:final2}
\rho_{S,\text{st.}}(y < 0) \approx  \L \sqrt{\D \tau} \int_0^{+\infty} \frac{\d w}{\sqrt{4 \pi}} \, (w + u_0) e^{-\frac{(w+u)^2}{4}} = 
\L \sqrt{\D \tau} \left[\frac{e^{-u^2/4}}{\sqrt{\pi}}+ \frac12 (u_0 -u) (1 - \text{Erf}(u/2))\right].
\ee
From this expression, the total volume $v^*$ exchanged during each auction is found to be:
\be
v^* = \int_{-\infty}^0 \d y \rho_{S,\text{st.}}(y) = \L {\D \tau} \left[\frac12 + \frac{u_0}{\sqrt{\pi}} \right] \approx 0.965 \L {\D \tau},
\ee
whereas the exact result (that can be obtained directly from the diffusion equation in the $\tau \to 0$ limit) is $v^* = \L {\D \tau}$. The error induced by our 
simple affine approximation is thus only a few percents. Interestingly, one sees that the total transacted volume $V$ in a finite time interval $T$, given by $V=v^* T/\tau$, 
remains finite when $\tau \to 0$, and equal to $V= \L \D T$. This observation should be put in perspective with the recent evolution of financial markets, where the time between transactions 
$\tau$ has become very small, while the volume of each transaction has simultaneously decreased, in such a way that the daily volume has remained roughly constant.

\subsection{The vanishing liquidity limit: From linear to square-root impact}

From the shape of the MSD close to transaction price given by Eq. (\ref{eq:final}), it is immediate to compute the supply and demand curves just before an auction when the inter-auction time $\tau$ 
tends to $0$. Denoting again as $y$ the difference between the price level $p$ and the fundamental price $\widehat p_t$, one finds: 
\be\label{eq:SD-final}
\begin{aligned}
S(p \geq \widehat p_t) &=  \L (y_0 y + \frac12 y^2) + v^*\\
D(p \leq \widehat p_t) &=  \L (-y_0 y + \frac12 y^2) + v^*
\end{aligned}
\ee
where, as found in the previous section, $y_0 \equiv u_0 \sqrt{\D \tau} \approx 0.824 \sqrt{\D \tau}$. {From Eq. (\ref{eq:final2}) above, it is readily seen that the supply (resp. demand) curve below (resp. above) $\widehat p_t$
can be written as $v^* F(y/\sqrt{\D \tau})$, where $F(u)$ is a certain function that goes from $F(0)=1$ to $F(\infty)=0$.}\footnote{This function reads, explicitly:
\be
\left[\frac12 + \frac{u_0}{\sqrt{\pi}} \right] F(u) =  \frac12 (1 - \text{Erf}(u/2))(\frac{u^2}{2} - u_0 u + 1) - \frac{e^{-u^2/4}}{\sqrt{\pi}} (\frac{u}{2}-u_0).
\ee
}

The above equation Eq. (\ref{eq:SD-final}) immediately allows us to compute the impact ${\cal I}(Q)\equiv  y^*$ of an extra buy quantity $Q$, as the solution of $\L (y_0 y^* + \frac12 y^{*2}) + v^* = Q + v^* F(y^*/\sqrt{\D \tau})$. 
It is clear that the solution can be written as $y^* = \sqrt{\D \tau} Y(Q/\L \D \tau)$, where $Y(q)$ obeys $u_0 Y + \frac12 Y^2 + (1 - F(Y)) = q.$ The limiting behaviours of $Y$ in the limits $q \ll 1$ and $q \gg 1$ are easy to compute, and read:
\be
Y(q) \approx_{q \ll 1} 0.555 q; \qquad Y(q) \approx_{q \gg 1} \sqrt{2q}.
\ee

 \begin{figure}[t!]
    \begin{center}
        \includegraphics[height=7cm]{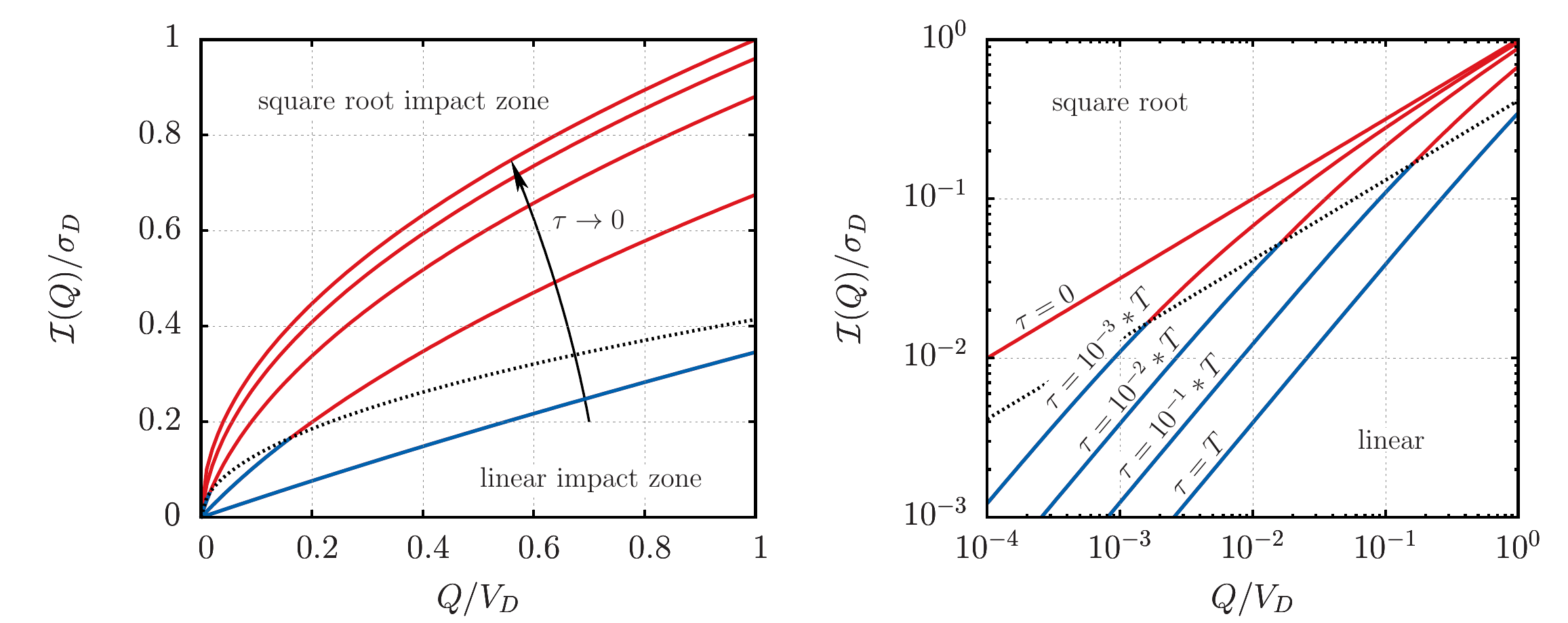}
    \end{center}
    \caption{The impact of traded volumes $Q$ for a given inter-auction time $\tau$ is linear for $Q \ll v^* = \L \D \tau$ and then square-root for $Q \gg v^* = \L \D \tau$. The linear impact zone shrinks to zero when $\tau \to 0$, when one recovers a pure square-root impact, i.e. a diverging Kyle's $\lambda$.}
    \label{fig:impacts}
\end{figure}

One therefore deduces that the impact ${\cal I}(Q)$ is linear in a region where the volume $Q$ is much smaller than $v^* \sim \L {\D \tau}$, i.e. when the extra volume is small compared to the typical volume 
exchanged during auctions, as expected. In the other limit, however, one recovers the {\it square-root impact} observed empirically (as found in \cite{donier2014fully}\footnote{In that paper, the study of market impact in the $\tau \to 0$ limit has been investigated much more in depth, in particular in the case of a progressive execution over some finite time window.}) :
\be
{\cal I}(Q \gg v^*) \approx \sqrt{\frac{2Q}{\L}},
\ee

The impact in the universal small $Q$ region\footnote{Corresponding to the region where $\phi_{\infty}$ is a good approximation of the order book, see Eq. \ref{eq:iteration}. For very large $Q$'s, the linear approximation describing the MSD curves breaks down, and one enters a presumably non-universal regime that is beyond the scope of the present discussion (cf. e.g. the shape of the MSD on the Bitcoin, Fig. \ref{fig:avg_book}).}, with a linear regime for $Q<v^*$ and a crossover to a square root regime when $Q$ becomes greater that $v^*$, is shown in Fig. ~\ref{fig:impacts}. Clearly, for $\tau = 0^+$, the auction volume $v^* = \L {\D \tau}$ also tends to zero, so that the region where impact is linear in volume shrinks to zero. In other words, when the interaction time becomes infinitely small, the impact of small trades is {\it never} linear. This comes from the fact that the MSD curves tend to zero exactly at the trading price when $\tau \to 0$.

\subsection{Empirical confirmation: the shape of supply and demand on the Bitcoin}

\begin{figure}[h!]
    \begin{center}
        \includegraphics[height=12cm]{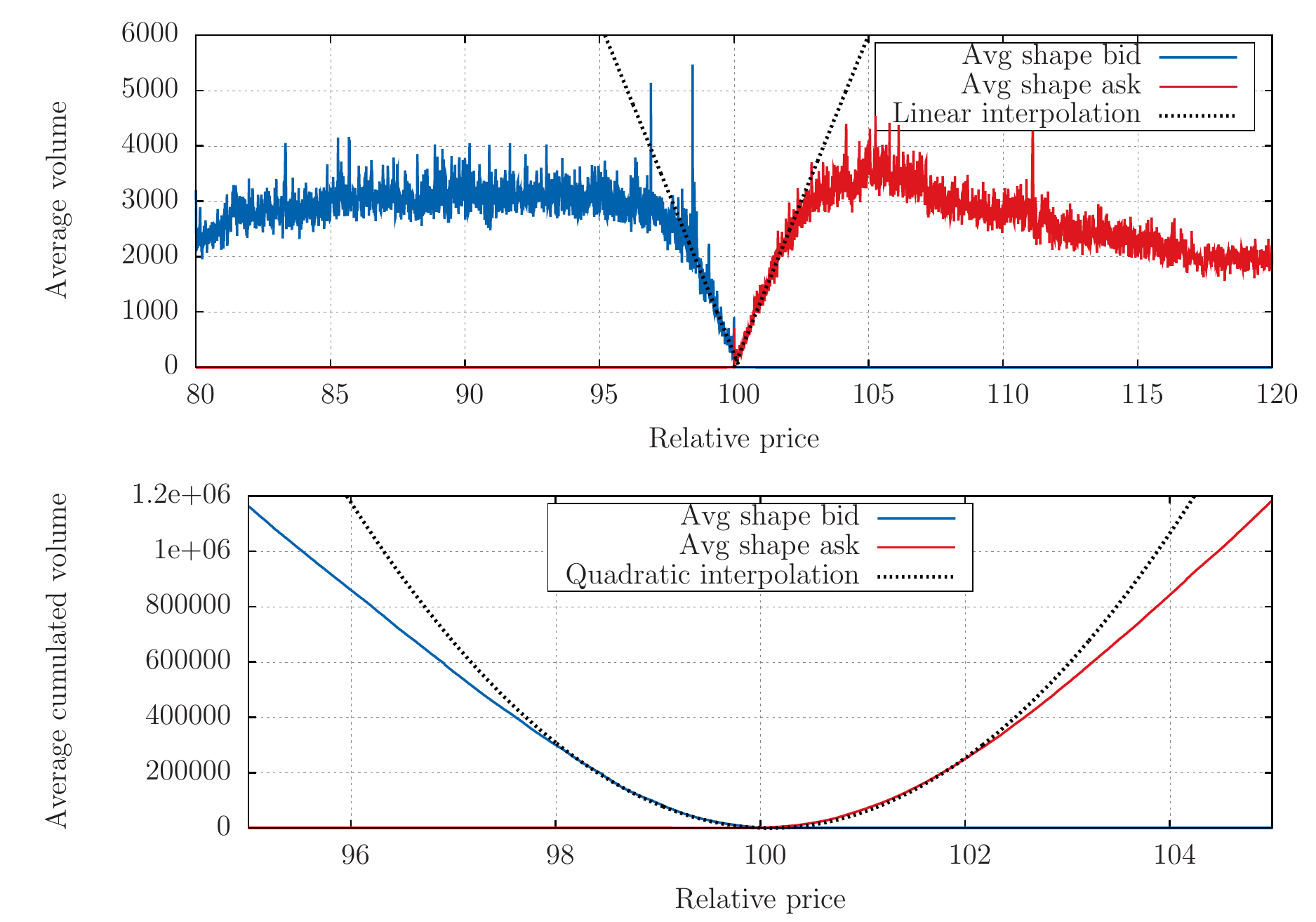}
    \end{center}
    \caption{Top: The average shape of the limit order book on the Bitcoin market, which we believe to be a representative sample of the MSD curve, in particular close to the price. The data comes from successive 
    snapshots of the full order book of the Bitcoin market, every 15 minutes from May 2013 to September 2013, centred auround the current mid-price. Bottom: 
    Integrated shape of the visible order book, as a proxy for the supply and demand curves on the Bitcoin. 
    The LOB/MSD curves grow linearly with respect to the distance to the price, resulting in a quadratic shape for supply and demand.}
    \label{fig:avg_book}
\end{figure}

Remarkably, we can check directly the above prediction on the shape of the MSD curves using Bitcoin data, where traders are much less strategic than in more mature financial markets 
and display their orders in the visible order book even quite far from the current price. The graph of the average shape of the Bitcoin limit order book (LOB) and of its integral, that are assumed to be good proxies for the MSD/SD curves at least not too far from the price, is shown on Fig. \ref{fig:avg_book}. 

Quite strikingly, the MSD curves are indeed \emph{linear} in the vicinity of the price that corresponds to about $5\%$ range, in perfect agreement with our dynamical theory of supply and demand in the limit of frequent auctions
(note in particular that $\partial_y S(p^*) = \partial_y D(p^*) \approx 0$!). Correspondingly, we do expect that impact of meta-orders should be well accounted by a square-root law in this region, which is indeed also found empirically 
(see  \cite{donier2014million} for the special case of Bitcoin case, and  \cite{Barra:1997,grinold2000active,Almgren:2005,Moro:2009,Toth:2011,Iacopo:2013,Gomes:2013,Bershova2013,Brokmann:2014,bacry2014market} for more mainstream markets). Further away from the price, the non-universal region clearly appears, where the shape of the MSD (here, approximately saturating to some constant value) depends on the detailed characteristics of the order flow, modelled in this paper 
by the functions $\nu(y)$ and $\omega(y)$.

\subsection{Summary}

\begin{figure}[h!]
    \begin{center}
        \includegraphics[height=10cm]{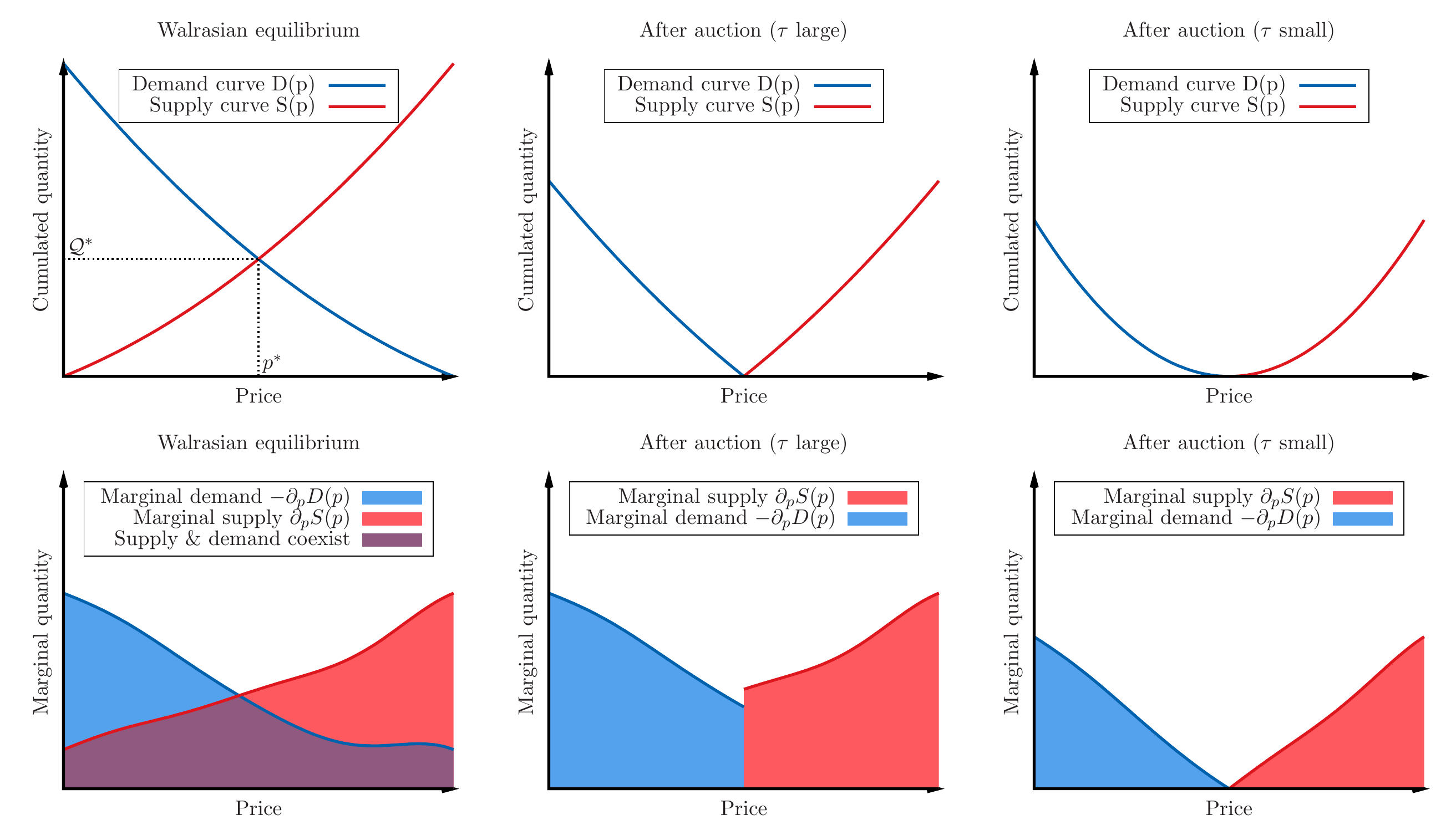}
    \end{center}
    \caption{Top: Supply and demand curves in (left) Walrasian auctions, (center) immediately after infrequent auctions and (right) immediately after frequent auctions. Bottom: Corresponding MSD curves. When transactions occur, supply and demand cannot cross (center and right). When the market is cleared frequently, supply and demand are depleted close to the price and exhibit a characteristic V-shape (right).}
    \label{fig:scheme}
\end{figure}

The above section presented our story in mathematical terms. The punchline is however quite simple, and well summarized by the graphs plotted in Fig. \ref{fig:scheme}, where we show (a) the standard Walrasian 
supply and demand curves just before the auction, from which the equilibrium price $p^*$ can be deduced; (b) the supply and demand curves just after an auction, when the inter-auction time $\tau$ is large enough -- in 
which case the marginal supply and demand are both finite at $p^*$; and (c) the supply and demand curves in the continuous time limit $\tau \to 0$, for which the marginal supply and demand curves vanish linearly around the 
current price, as found in the Bitcoin market. 

\section{Discussion}\label{sec:discussion}

Up to now, our presentation has been fairly technical, with the aim of establishing our main results. Still, many points of general interest have 
glossed over for the sake of readability. We feel that these deserve a more detailed discussion that we provide now.  

\subsection{Price discovery vs. price formation}

The interpretation of price moves in financial markets has generated endless theoretical debates, culminating in the split 2013 Nobel prize between the 
hero of efficient market theory (\cite{fama1970efficient}), and the beacon of behavioural economics (\cite{shiller1980stock}). For the former school of thought, the \emph{fundamental price} 
of an asset pre-exists and only waits to be ``discovered'' by aggregating the unbiased opinions of rational market participants. Up to small self-correcting errors,
markets do clear at the right price. This assumption is a pre-requisite for many economic models, e.g. the classical Kyle model  \cite{kyle1985continuous} within which some agents are \emph{informed}: 
they are supposed to know the fundamental price for the next period and invest accordingly so as to make a maximum profit from of this information. 

This perfect Platonian view of markets is however hard to swallow for many followers of Shiller. After all, financial markets are 
driven by humans who have a very imperfect knowledge of the fundamental price, prone to many behavioural biases.  The market is a device 
that merely aggregates all participants' intentions, 
regardless on whether they are justified or not, and spits out the ``market price''. In this view, markets allow \emph{price formation} rather than \emph{price discovery}. 

Our modelling strategy provides a very intuitive framework to think about the difference between the two viewpoints. In Sec. ~\ref{sec:behaviour}, we postulated that on any time interval $(t, t+{\rm d} t)$ 
an information item is released, leading to a change in the (unobservable) fundamental price ${\rm d} \xi_t$, of variance $\sigma^2 \d t$. This piece of news is however 
interpreted by agent $i$ as predicting a price change $\beta_t^i {\rm d} \xi_t$ plus idiosyncratic noise. As noted above, this means that the news is over-interpreted by some 
($\beta^i>1$) and under-interpreted by others ($\beta^i<1$). If for any news ${\rm d} \xi_t$, the $\beta_t^i$'s average \emph{exactly} to 1, the market on aggregate perfectly digests the news and the (permanent) increment in 
perceived price is precisely ${\rm d} \xi_t$. If however this is not the case, $\E_i[\beta_t^i] {\rm d} \xi_t  \neq {\rm d} \xi_t$, and this leads to two possibly different definitions of a reference price: the 
{\it fundamental} price $p_t^{\text{F}}$ (unknown to agents) and the {\it market} price $\widehat{p}_t$ (encoding agents perceptions):\footnote{Here and below, we assume that agents perceptions are symmetrically 
distributed around $\widehat{p}_t$, that we identify with the auction clearing price $p^*$. In other words, we neglect here any ``bid-ask'' bounce that may affect the short-time price dynamics.}
\be\label{eq:prices}
\begin{aligned}
&p_t^{\text{F}}\equiv  p_0 + \int_0^t {\rm d} \xi_s\\
&\widehat{p}_t \equiv  p_0 + \int_0^t \E_i[\beta_s^i] {\rm d} \xi_s ,
\end{aligned}
\ee
where $p_0$ is an arbitrary reference price, assuming that the market price is equal to the fundamental price initially. Only if the average market reaction $\beta_t\equiv \E_i[\beta_t^i]$ is unbiased at any time can one speak 
of efficient markets and \emph{price discovery}. Otherwise, \emph{price formation} prevails, and the market price errs away from the fundamental price, as envisaged by  \cite{black1986noise}. More precisely, one can compute the 
pricing error as:
\be
\text{Var}(\widehat{p}_t-p_t^{\text{F}}) = \mathbb{E}\left[\sigma^2 \int_0^t (\beta_s-1)^2 {\rm d} s\right] = \text{Var}(\beta_t)\text{Var}(p_t^{\text{F}}), 
\ee
as well as the variance of the market price as: 
\be
\text{Var}(\widehat{p}_t) = \mathbb{E}\left[\sigma^2 \int_0^t \beta_s^2 {\rm d} s\right] = \left[1+\text{Var}(\beta_t)\right] \text{Var}(p_t^{\text{F}})
\ee
showing that even if there is no bias on average (i.e. $\E_{s}[\beta_s] \equiv 1$), $\text{Var}(\widehat{p}_t) \geq \text{Var}(p_t^{\text{F}})$ with a strict inequality as soon as the market reaction is not perfectly 
unbiased \emph{at all times} -- a highly plausible situation.\footnote{Note however that on long time scales, some weak mean reversion towards the fundamental price should take place, as discussed in e.g. \cite{black1986noise,de1990security,bouchaud1998langevin}.} 
This embeds Shiller's famous excess volatility puzzle  \cite{shiller1980stock} in an interesting formal framework.

We are here at the core of a crucial question in financial economics: what is ``information''?  The discussion above naturally leads to two definitions of information: 
\begin{enumerate}[(a)]
\item information on the \emph{fundamental price} (often called fundamental information), corresponds to some (perhaps noisy) knowledge about the value of $p_t^{\text{F}}$, while 
\item information on the \emph{market price}, corresponds to some (perhaps noisy) knowledge about the future value of the market price $\widehat{p}_t$. 
\end{enumerate}

In the latter case, information is only about \emph{correctly anticipating the behaviour of others}, exactly as Keynes envisioned  \cite{keynes2006general}. 
The notion of \emph{information} should then rather be replaced by the notion of \emph{correlation} -- if all the market participants' $\beta$ were negative, the correct information for an arbitrageur would correspond to also interpreting the news with a negative $\beta$, even if it did not make sense. In this context, the difference between a ``noise'' trader and an ``informed'' trader in Kyle's model is merely the level of 
correlation with the crowd of other market participants: informed traders are positively correlated with it, whereas noise traders are simply uncorrelated with the crowd \footnote{See the detailed discussion in \cite{donier2014fully}, 
and \cite{Gomes:2013,donier2014million} for empirical studies on how ``informed" and ``noise" trades impact the price differently on the long run. While both have similar instantaneous impacts, on the long run the price reverts 
to its initial value for noise trades whereas a permanent component remains for ``informed'' trades. Interestingly, data suggests that the permanent impact is of the same order of magnitude as the transient impact.}. 

Finally, let us give an alternative interpretation of the market price equation Eq. (\ref{eq:prices}), in terms of the individual estimate of the fundamental price $\widehat p_t^i$ and the fraction of market shares $F^i$ of agent $i\in\{1,\dots N\}$, such that $\sum_i F^i=1$. The individual estimate of the fundamental price evolves as ${\rm d}\widehat{p}_t^i = {\rm d} \xi_t^i$. Assuming that expectations are symmetrically distributed around $\widehat{p}_t$, the market clearing price $\widehat{p}_t$ is given by the weighted average of individual prices (whether they are truly ``informed'' or not):
\be
\widehat{p}_t = \sum_i F^i \widehat{p}_t^i, 
\ee
which of course coincides with Eq. (\ref{eq:prices}). However, one now clearly sees that the \emph{permanent} impact of an agent with market share $F^i$ on the price is $F^i {\rm d} \xi^i$ when his view on the price changes by 
${\rm d}\xi^i$. When the market share of an individual investor is small, the permanent impact of his/her isolated orders on market prices is itself small (in particular much smaller than the transient square-root impact that  depends on his/her fraction of the daily volume), except if his/her trade is correlated with the rest of the market, in which case it is often said to have an \emph{alpha} -- i.e. a predictive signal on the price (see previous footnote).

\subsection{Market stability and marked-to-market valuation} 

The bottom line of the model and analysis presented in the above sections is the possible effect of market design itself on \emph{price stability}. Indeed, the highly singular square root impact consistently measured on financial markets implies that small perturbation (e.g. noise trading) may result in abnormally high returns, questioning the robustness of the price formation and the very stability of markets. Indeed, if one freak order, whose volume 
remains small in comparison to market daily volume, can move the price by a large fraction of the daily volatility, then it is clear that one should not put too much faith in the reliability and resiliency of market prices. 
As we argued above, this market fragility is the result of continuous market clearing, that leads to the following property:
\begin{center}
The price is the point at which a {\it vanishing} supply meets a {\it vanishing} demand.\\
\end{center}
An implication of particular importance is the relevance of mark-to-market accounting rules for the valuation of large portfolios (so as to leave aside the problem of liquidation costs that usually enters in the discussion, one can think of the assets of an insurance firm that are kept until expiry). From the previous discussion, it is meaningless -- or even dangerous -- to mark too closely portfolios to the market prices, as some noise traders, fat-fingers or even ill-intentioned manipulators can trigger large re-balancing with not-so-large volumes, resulting in inappropriate profits and in unstable prices. 

It would be more significant to assess market prices on the basis of the local supply and demand (around the market price) so as to smooth out fluctuations. A practical way to do so is to monitor the market \emph{liquidity} $\cal L$, that is well proxied by the ratio $\sigma_D/\sqrt{V_D}$ (where $\sigma_D$ is the daily volatility and $V_D$ is the daily traded volume), as proposed and tested on the Bitcoin market in  \cite{donier2015markets}. Several other measures of liquidity might also be relevant, see \cite{amihud1986asset,foucault2013market,corradi2015liquidity}. In the case when the asset is not assumed to be kept until expiry but might incur liquidation costs, the monitoring of liquidity would be a good indicator of the \emph{ex-post} value of the portfolio, once liquidated. This is the idea of impact-discounted mark-to-market value proposed and discussed in  \cite{caccioli2012impact}.

\subsection{Would batch auctions be beneficial?}

In order to curb potentially nefarious and socially wasteful HFT activities, a possibility that is currently hotly debated  \cite{budish2013high} is to change the continuous trading system to frequent batch auctions, that would occur in discrete time with a time interval between auctions of order $100$ms-$1$s.  While we are not necessarily convinced that high frequency trading is such an evil\footnote{The overall profit of HFT firms was estimated 
to be around 5B\$/year at its peak in 2010. This corresponds to an estimated cost of 1 basis point ($10^{-4}$) per transaction in the US equity markets, 
compatible with a fraction of the average bid-ask spread in the same period  \cite{hendershott2011does}. This is probably at least 10 times smaller than the profits of ``old school" market makers: the average bid-ask spread on US markets was fluctuating around 60 basis points (!) from 1900 to 1980, before declining sharply  \cite{jones2002century}.} and the investment in speed technologies is such a waste on the long run, we believe that the issue of market stability should be of primary concern. Our theory provides a natural framework for studying the effect of such 
market design changes on the supply and demand curves, and on the stability of the resulting price.

From the above analysis, we concluded that the singularity at the immediate vicinity of the price is regularized when auctions occur in discrete time $\tau$, leading to a reduced price impact (linear instead of square-root). 
However, one should not rejoice too fast, since we also saw that this regularisation only concerns very small volumes, less than the average volume traded in the market during time $\tau$. In order to have a substantial 
effect and reduce the impact of trade size $Q$ of -- say -- ${ 1-}10$\% of the average daily volume, the inter-auction time should be, unsurprisingly, of the order of five minutes to one hour and not milliseconds or even seconds. Trying to improve market stability through frequent batch auctions does not seem very useful in the light of our theory, except if one accepts to clear the market every hour or so, which would potentially imply other problems, 
such as and new source of liquidity risk and the corresponding development of secondary markets where transactions would take place in-between auctions.  

\section{Conclusion}\label{sec:conclusion}

In this paper, we have developed a fully dynamic theory of liquidity, based on weak and general assumptions on investors' behaviours: in a nutshell, heterogeneous reactions to incoming news of a large number of ``infinitesimal''
investors. Addressing the inability of 
the Walrasian theory to take transactions into consideration, we allow for auctions to clear the market periodically, and show how the market clearing mechanism itself affects the structural properties of supply and demand. In the case when the time between auctions is very large, we recover classical Walrasian auctions, in which market prices and liquidity are determined by the long-term (im-)balance between the incoming supply and the incoming demand. When auctions are allowed to happen at high frequency, the liquidity around the price mechanically vanishes, which leads to an anomalous, square-root impact of small orders that increases with market \emph{volatility}. This 
accounts for the universally observed square root impact of small orders on modern financial markets and on the Bitcoin. In order to obtain a direct confirmation of the theory, we measured the shape of the Bitcoin order book (that appears to be a faithful reflection of low frequency supply and demand, at least close enough to the price), which indeed displays a striking ``V-shape'' for the marginal supply and demand (see Fig. \ref{fig:avg_book}). 

Our results highlight an apparent paradox: the more frequent the transactions are allowed to occur (thereby increasing, in theory, market efficiency), the more {\it fragile} the resulting price is! In continuous double auctions
markets, the price can be seen as the point at which a \emph{vanishing} supply meets a \emph{vanishing} demand, challenging the Platonian view of financial markets that prices are well-defined and stable. Our framework allows us to draw two further conclusions of general interest. First, the local estimates of supply and demand needs to be taken into account when one watches market prices, with important implications on portfolio valuation and stability monitoring \cite{kyle2012large,donier2015markets}. Second, as soon as the reaction to incoming news is not unbiased at all times, the volatility of the market price exceeds the fundamental volatility, embedding Shiller's famous excess volatility puzzle \cite{shiller1980stock} in an interesting formal framework.

Although we only considered market design and price stability within a particular angle, we believe that our framework can be extended to address many other practical problems, such as the market maker's problem, cross-impact between markets, optimal execution issues or the effect of taxation and other changes in market design. We also restricted to stationary market conditions, but our formalism can be readily adapted to include fluctuations in liquidity and/or market volatility (that would translate in some stochastic evolution of ${\cal D}$). Finally, the surprisingly good agreement between theory and empirical data suggests to extend our set of hypotheses to agent-based models of markets (cf Sec. \ref{sec:kyle}), with the aim of producing realistic emergent properties from microscopic agent behaviour. More generally, our results vindicate an approach to dynamics in economic sciences as resulting from complex interactions between many heterogeneous agents, that may -- or may not -- be rational, in the general vein of, e.g. \cite{hommes2006heterogeneous,gualdi2015tipping} and refs. therein. The resulting partial differential equation that governs the evolution of agents' preferences offers a much deeper level of understanding than simply postulating ad-hoc stochastic models for prices. We believe that this modelling strategy can be extended to many other situations of economic relevance. 

\section*{Acknowledgements}

We warmly thank M. Abeille, J. Bonart, R. Cont, J. De Lataillade,  D. Delli Gatti, M. Gould, T. Hendershott, J. Kockelkoren, C. A. Lehalle, Y. Lemp\'eri\`ere, I. Mastromatteo, M. Potters, I. Rosu and B. T\'oth for many crucial discussions 
and collaborations on these issues -- and in particular G. Zerah for his help on the Wiener-Hopf method. We also thank A. Tilloy for useful remarks on the manuscript.  

\bibliographystyle{plainnat}
\bibpunct{(}{)}{;}{a}{,}{,}
\bibliography{biblio}

\end{document}